# Average energy required to produce an ion pair, revisited

E. Fourkal[1], A. Nahum[2], C-M. Ma[1], K. Paskalev[1]

[1]Radiation Oncology Dept., Fox Chase Cancer Center, 333 Cottman Ave., Philadelphia 19111,USA,

[2]Physics Department, Clatterbridge Centre for Oncology, Clatterbridge Road, Merseyside CH63 4JY, UK

ABSTRACT

The present work is a theoretical/computer simulation study aimed at investigating the dependence of the $w$ value in air on the initial energy of the ionizing electrons. The current assumption of a constant $w$ value underpins the accurate determination of absorbed dose in megavoltage radiotherapy. The transport of electrons in the energy range between 1 keV and 10 MeV has been studied using the EGSnrc Monte Carlo code system. We have extended the electron degradation spectra calculations down to 200 eV by solving an integral equation derived from the cavity integral formulation for the absorbed dose. The present study confirms the constancy of the $w$ value in the megavoltage energy range though our calculation method is not capable of accurately predicting its absolute value. When the average electron energy falls below several keV, it is well known that $w$ exhibits an increase in its absolute value which our calculations also indicate.

## 1. Introduction

As electrons traverse the matter they lose kinetic energy in ionization and excitation events as well as being subject to elastic scattering by the atoms/molecules of the medium. Primary electrons generate many secondary particles of varying energies as a result of ionization collisions and this results in specific energy distributions (so-called 'slowing-down' or 'degradation' spectra) that depend on the initial energy of the ionizing electrons and the composition and geometry of the scattering medium. It is important to be able to predict these distributions since they implicitly determine the ion yield or the total number of ion pairs produced. The knowledge of this function will yield the differential $w=\Delta E/\Delta N$ value, which is the mean energy necessary to produce an ion pair, where $\Delta E$ is the mean energy lost by a charged particle of energy $E$ and $\Delta N$ is the mean number of ion pairs produced as a result of the given energy absorption. At this point we would like to emphasize the difference between the differential $w$ value, which is under investigation in this work and the $W$ value defined as a quotient of by $N$, where $N$ is the mean number of ion pairs formed when the initial kinetic energy $E$ of a charged particle (electron) is completely dissipated in a gas. These two quantities are related through the following expression

$$W(E) = \frac{E}{\int_{I_0}^{E} \frac{d\varepsilon}{w(\varepsilon)}} \tag{1}$$

where $I_0$ is the ionization threshold.

The majority of the experimental investigations have been directed toward measurement of the $W$ value, which is only possible for low electron energies, because of the requirement of complete dissipation of the initial electron energy in air. For megavoltage electron beams however, such measurements would require calorimetric systems with depth dimensions on the order of several tens of meters, which would constitute a formidable experimental challenge. To the best of our knowledge there have not been any direct experimental investigations (i.e. calorimetric measurements in air not in graphite or in another medium) of the $w$ value for megavoltage electrons. The exact knowledge of $w$ is of paramount importance for radiation dosimetry of megavoltage electron (and photon) beams since it allows one to determine the energy absorbed in the air volume occupied by the ion chamber (radiation dose at a given spatial location) through the measurement of the charge collected by the same chamber.

Ion chambers play a central role in radiotherapy dose determination. It is implicitly assumed that they behave as Bragg-Gray cavities[1] and therefore that their response is proportional to the medium-to-air stopping-power ratio (multiplied by any minor perturbation factors e.g. $p_{cav}$, $p_{cel}$). This implies that the dose to the air in the cavity, $D_{air}$, suitably corrected for any recombination, polarity effects and temperature and pressure, is proportional to signal from the ion chamber. As is well known this $D_{air}$ is derived from[2]

$$D_{air} = J_g \frac{w}{e} \tag{2}$$

where $J_g$ is the ionization per unit mass of a gas (the electrometer reading can be assumed to be proportional to this quantity) and $(w/e)_{air}$ is the mean energy required to create an ion pair in air. The absorbed dose to the medium is generally written in the following form

$$D_{med} = D_{air} s_{med,air} \prod_i p_i \qquad (3)$$

and all the subsequent radiation-quality dependent behavior of the ionization chamber is assumed to reside in the factors $s_{med,air}$ and $p_i$. We will not be concerned in this study with the details of the various sources of perturbation, nor in fact with the details of the evaluation of $s_{med,air}$.

The key assumption underlying all the above is that any dependence of $(w/e)_{air}$ on beam quality is entirely negligible. However, there have been suggestions that $(w/e)_{air}$ may vary with beam quality.[3] Furthermore, there are some theoretical grounds for believing that such a variation cannot be ruled out. Using simple classical-physics reasoning, which relates the energy transferred to the medium $\Delta E \sim 1/b^2$ to the impact parameter $b^4$ and the fact that at the relativistic energies the impact parameter increases by the relativistic factor $\gamma$, it follows that the increase in collision stopping power corresponding to 'distant collisions' via the increase in $b$ consists of a greater proportion of excitations over ionizations (because of small energy transfers) than is the case at sub-relativistic energies. This in turn might be expected to influence the dependence of $(w/e)_{air}$ on the electron energy in the relativistic domain, specifically resulting in a gradual increase in $(w/e)_{air}$ as the ionizing electron energy increases. This reasoning however, relies heavily on the classical depiction in which the projectile slows down continuously, giving up its energy to the medium in transfers that could be smaller than that corresponding to the lowest possible atomic excitation. In reality, the individual acts of energy transfer are described within the realm of the quantum theory and the notion of *continuous* energy deposition should only be understood in a statistical sense in which, on the average over many collisions, a small energy is transferred (classical picture) which is equivalent to appreciable amounts of energy (or quanta) transferred in a very small fraction of those collisions.[5] Therefore, any reliable calculations of $w/e$ have to intrinsically depend on the quantum mechanical description rather than on the classical reasoning.

Is there any experimental evidence that $(w/e)_{air}$ varies with radiation quality? This is a difficult question to answer definitively. In principle it could be addressed by comparing a relative depth-dose curve determined in an electron beam using an ionization chamber against one determined by a completely independent method such as calorimetry, or with some other detector with an extremely well-known energy response. In practice there are almost no detectors which fulfill this precondition at the sub 1% (or preferably sub 0.3%) accuracy required. Another possible method of determining a (relative) depth-dose curve is by Monte-Carlo simulation. However, this requires an exquisitely detailed model of the treatment head of the linear accelerator as well as an absolute method of determining the energy (and possibly the energy spectrum) of the electrons emerging from the vacuum window of the accelerator waveguide system.[6] A further problem with this method is the (admittedly) small difference between depth-dose curves derived from different Monte-Carlo codes[7] or from the same code, e.g. EGSnrc, using slightly different assumptions about the electron physics e.g. the exact cross sections employed for electron single/multiple scattering and the key material-dependent parameter in the Bethe-Bloch formula for the stopping power, the mean excitation energy *I*. Summarizing the above, at the present time it is not possible to pin down the energy independence of $(w/e)_{air}$ to better than $\approx 1\%$ over the 1 to 20-MeV energy range in electron beams.

The quantitative prediction of the $w$ value began with the celebrated work of Bethe.[8] Although extensive efforts have been made since then,[9] the subject still remains incompletely understood. Any theory of $w$ must answer a number of questions concerning the magnitude of the $w$ value. First, it should be able to explain the absolute value for the given particle modality. Second, it should explain why $w$ does not depend on the energy of the ionizing particle assuming that this energy is larger than the energy of the valence electrons of the molecule. Third, when the energy of ionizing particle is comparable to the energy of the outer shell electrons, it should explain why $w$ becomes dependent on the ionizing particle energy. The general trend is that the $w$ increases as the velocity of the ionizing particle decreases in this very low energy region.

As has been mentioned above, the measurement of $w$ requires the determination of both the mean number of ion pairs and the mean energy imparted to the absorbing material.[10,11] The measurement of the number of ion pairs usually encompasses the determination of the total ionization current or total charge collected over a given period. In most experiments the ionization measurements are performed with cavity chambers and the energy absorbed is obtained from the calorimetry, since most of the kinetic energy of electrons lost in an absorber (air plus chamber material) appears as heat (one also has to account for the small fraction of kinetic energy transformed into chemical energy or escaping in the form of secondary radiation). The International Commission on Radiation Units and Measurements[9] published a compilation of available experimental values for $w$ for electrons, and a mean value of 33.85 eV was recommended for dry air.

There have also been a number of theoretical/computer simulation studies of low-energy electron transport in air.[12,13, 14,15] Theoretically obtained $w$ values of Grosswendt *et al.*[12] for primary electron energies in the range between 50 eV and 5 keV closely follow those experimentally measured by Waibel *et al.*[10] The authors used a Monte Carlo method to simulate the trajectories of electrons directly from elastic and inelastic cross section data without resorting to the continuous-slowing-down approximation and multiple scattering theories. Even though they were able to predict both the shape of the $w$ dependence on the initial electron energy and its magnitude, some questions arise from the authors' treatment of this issue. We will consider these questions in subsequent sections of this work.

The main purpose of this paper is to examine in detail the assumption of the absolute constancy of $(w/e)_{air}$ at megavoltage radiation qualities. This will involve describing the electron transport related to the analysis of ionization chamber measurements with particles in the energy range between 200 eV and 10 MeV with a special emphasis on the calculation of the $w$ value. The relative composition of air is assumed to be 78 % $N_2$ and 22 % $O_2$ with negligible contribution of Argon and other chemical compounds. Since we are predominately interested in the energy dependence of $w/e$ for the megavoltage beams, the well known Jesse effect[16] (more profound for low energy particles) is not considered in the present work. We have also neglected the contribution of Auger electrons in the electron fluence calculations. As shown in Ref. ([17]) their contribution to the total electron fluence in the sub keV energy range is small.

The main bulk of calculations is based on the EGSnrc Monte Carlo code.[18] EGSnrc system as its predecessor EGS4 uses the so-called condensed history technique in which many track segments of a real random walk are grouped together into a single step.[19,20] The cumulative effects of inelastic collisions with energy loss less than the cutoff energy for discrete $\delta$–ray production[1] are taken into account by sampling energy and directional changes from appropriate multiple scattering distributions.

We wish to stress that our aim is not a mere prediction of $w$ value for a wide range of electron energies, but rather a quantitative understanding of the issues raised earlier in this section with specific emphasis on the role played by different microscopic processes

---

[1] This cutoff is denoted by AE in the EGSnrc code system.

in the observed value for *w*.

# 2. Methods

## 2.1. Electron degradation spectra and ion yield calculations

As the ionizing electron traverses the medium it interacts with molecules/atoms of this medium via both elastic (i.e. no energy transfer) collisions, producing only changes in direction, and inelastic collisions, which result in the transfer of energy from the ionizing particle to the medium. In some instances, the transferred energy goes into the creation of secondary 'free' electrons (ionization events) and in others the energy is absorbed by the molecules and distributed through their internal degrees of freedom. These latter inelastic processes do not lead to ionizations and are known as *excitation* collisions. As the primary electrons slow down they generate many secondary electrons of varying energies. These electrons are characterized by certain energy distributions that depend on the primary electron spectrum and the scattering medium. Following Spencer and Fano,[21] for electrons of initial energy $E$ we use $y(E, E')dE'$ to denote the total pathlength of all generations of electrons (primary, secondary, tertiary, etc.) having energies between $E'$ and $E' + dE'$. This function is the so-called degradation spectrum of electrons. It is important to note that the degradation spectrum is related to the electron fluence distribution differential in energy $d\Phi/dE$ through the following equation,

$$\frac{d\Phi}{dE'} = \frac{1}{V} y(E, E')$$

(4)

where $V$ is the volume of the region in which the spectrum is calculated. Spencer and Fano gave an integral equation for $y(E, E')$ and discussed methods of its numerical solution.

In present work we use Monte Carlo simulation to obtain $d\Phi/dE$ as a function of the initial primary electron energy and depth. The simulation geometry is shown in Figure (1), where a parallel beam of electrons of radius 1 cm is incident on a water phantom. An air slab of thickness 0.5 cm is positioned at different depths in the water phantom to simulate the ion chamber measurements. The FLURZnrc usercode, which is a part of EGSnrc Monte-Carlo system, was used to calculate the electron differential fluence distribution $d\Phi/dE$. Once the fluence distributions are known the degradation spectra can easily be obtained from equation (4).

The ion yield $N(E)$ can then be calculated as,

$$N(E) = N_m \int_I^E \sigma_{ion}(E') \frac{d\Phi}{dE'} dE' \qquad (5)$$

where $N_m$ is the number of molecules in the slab and $\sigma_{ion}(E)$ is the total or *gross* ionization cross section defined as

$$\sigma_{ion}(E) = \sum_m m\sigma_m(E) \qquad (6)$$

where $\sigma_m(E)$ is the cross section for producing an ion of *m*th degree of ionization in the single collision (*m* electrons are ejected from the atom/molecule). The summation is over all charge states (degrees of ionization) that contribute to the collected/measured charge/current. As one can see from equation (5), the exact knowledge of ionization cross sections is essential for the proper calculation of the number of the electron-ion pairs produced by the ionizing particle of energy $E$.

To calculate the electron impact ionization cross section for creation of singly charged ions (single electron is ejected from the atom/molecule) we have used the binary-encounter dipole model proposed by Kim and Rudd.[22,23] This model successfully combines the modified Møller theory, which describes the collision of an incident electron with bound electrons of an atom/molecule, accounting for *hard* collisions (collisions with relatively large energy transfers comparable to the energy of the ionizing particle) with the Bethe theory, which describes the *soft* or *distant* collisions with small energy transfers. This type of collision originates from a dipole type interaction between the incident particle and an atom/molecule and prevails at high energies of the ionizing electron (first-order perturbation theory calculations). The fusion of both theories leads to the following expression for the ionization cross section of a given molecular/atomic orbital,

$$\sigma_1 = \frac{4\pi a_0^2 \alpha^4 N}{(\beta_t^2 + \beta_u^2 + \beta_b^2)2b'} (\frac{1}{2}\left[ \ln\left( \frac{\beta_t^2}{1-\beta_t^2} \right) - \beta_t^2 - \ln(2b') \right]$$

$$\left( 1 - \frac{1}{t^2} \right) + 1 - \frac{1}{t} - \frac{\ln t}{(t+1)} \frac{1+2t'}{(1+t'/2)^2} + \frac{b'^2}{(1+t'/2)^2} \frac{t-1}{2}) \qquad (7)$$

where *N* is the orbital electron occupation number, $\alpha$ is a fine structure constant, *b'* is the orbital binding energy in the units of electron rest mass, *t'* is the kinetic energy of the incident electron in the units of electron rest mass, *t* is the kinetic energy of the incident electron in the units of the orbital binding energy, $\beta_t^2 = 1 - 1/(1+t'^2)$, $\beta_b^2 = 1 - 1/(1+b'^2)$, $\beta_u^2 = 1 - 1/(1 + u'^2)$ with u' being the average orbital kinetic energy of the target electrons in the units of electron rest mass (for a given orbital). The total ionization cross section for production of singly charged ion is obtained by summing equation (7) over all molecular orbitals. The values for orbital parameters of different chemical compounds including molecular oxygen and nitrogen can be found in the NIST Physical Reference database and references therein.[24]

As mentioned earlier, another possible ionization mechanism is the simultaneous removal of several (i.e. more than one) atomic/molecular electrons following the impact of the single electron on a neutral atom/molecule (single step multiple ionization); this can be described as

$$A + e \rightarrow A^{m+} + (m+1)e \qquad (8)$$

Calculations of multiple atomic/molecular ionization processes using a quantum mechanical description are extremely difficult for the majority of targets[25,26] since one has to consider two or more continuum electrons and their mutual interactions as they leave an atom/molecule. Experimental data for the formation of highly charged atomic/molecular ions are scarce for the most atoms because of the fact that the cross section decreases rapidly with increasing charge state of the final ion. Therefore one has to rely on semi-empirical and semi-classical approaches to determine multiple ionization cross sections. The semi-classical Deutsch-Märk (DM) formalism for the calculation of cross sections for the formation of multiply charged ions has been proposed.[27] The proposed cross section $\sigma_{m+}$ for the formation of an ion $A^{m+}$, which is generally expressed as the product of $m$ independent terms each describing the removal of a single electron, has the form

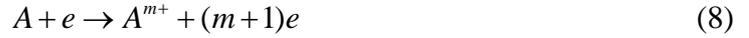

$$\sigma_{m+} = g^m \sum_k \pi (r_k)^2 \xi_k f_k(U) \qquad (9)$$

where the summation extends over the various atomic/molecular subshells with k=1 referring to the outermost subshell, etc., $(r_k)^2$ is the mean square radius of the atomic/molecular subshell, $\xi_k$ in the number of electrons in that subshell, and $g^m$ are weighting factors determined from a fitting procedure (for details, see the paper by Deutsch $et\ al.$[28]). The functions $f_k(U)$ describe the energy dependence of the ionization cross section,

$$f_k(U) = \frac{1}{U}\left[\frac{U-1}{U+1}\right]^a \left(b + c\left[1-(2U)^{-1}\right]\ln\left[2.7+(U-1)^{0.5}\right]\right)$$ (10)

where $U = E/E_m$, where $E$ is the energy of the incident electron and $E_m$ is the ionization energy required for the simultaneous removal of $m$ electrons from atom/molecule. The energy-dependent functions $f_k(U)$ are different for different contributing subshells and molecular/atomic species and their detailed discussion is given by Margreiter et al.[29] In their recent papers[30,31] the authors have applied this model for the calculation of multiple ionization cross sections for oxygen and nitrogen species. Combining their results with the binary-encounter model of Kim and Rudd using equation (6), we obtain the total electron impact ionization cross section for nitrogen and oxygen gas, shown in figures (2) and (3). The calculated total electron impact ionization cross sections for both oxygen and nitrogen species are in a good agreement with those compiled by Kieffer and Dunn.[32]

## 2.2. Cavity integral formulation of the absorbed dose and an expression for the w value

We have computed the energy absorbed in the medium (air) using both Monte Carlo simulation (i.e. scoring the energy deposited in a small volume) and through the Spencer-Attix formulation of the absorbed dose in an elementary volume[33] given by the following expression,

$$D = \int_\Delta^{E_0} L_m(E,\Delta)\frac{d\Phi}{dE}\,dE + S_m(\Delta)\frac{d\Phi(E)}{dE}\bigg|_{E=\Delta}\Delta$$ (11)

where $\Delta$ is the cutoff energy, $L_m(E,\Delta)$ is the restricted mass collision stopping power for electrons of energy E, which includes only energy losses not exceeding $\Delta$, $S_m(\Delta)$ is the collision mass stopping power for electrons of energy $\Delta$ and $\frac{d\Phi}{dE}(\Delta)$ is the differential energy fluence distribution evaluated at $\Delta$; it is the latter quantity that is furnished by the Monte-Carlo simulation The last term in equation (11) is the so-called track-end contribution to the absorbed dose, introduced by Nahum.[34] This term accounts for energy deposited by electrons with kinetic energies equal to the cutoff energy. The dose deposited by the electrons in the medium calculated using equation (11) should be identical to that calculated using DOSRZnrc Monte Carlo code. Combining equation (11) with (5) one arrives at the following expression for the $w$ value,

$$w_\Delta = \frac{\rho}{N_m} \frac{\int_\Delta^{E_0} L_m(E',\Delta) \frac{d\Phi}{dE'} dE' + S_m(\Delta) \frac{d\Phi(E)}{dE}\Big|_{E=\Delta} \Delta}{\int_\Delta^{E_0} \frac{d\Phi(E')}{dE'} \sigma_{ion}(E') dE'}$$

(12)

where $\rho$ is the density of the medium (air). The subscript $\Delta$ in Eq. (12) makes explicit the dependence of this theoretically obtained $w$ on the cutoff energy. The higher the cutoff energy $\Delta$ the larger the $w$ value will be as the true number of ions will be underestimated. Of course the experimentally measured $w$ value corresponds to the cutoff energy being equal to the ionization threshold, which is on the order of 15 eV for air.

The calculations performed using the FLURZnrc simulation code give differential electron fluence distributions in the range between a cutoff energy chosen by the user but subject to the limits inherent in the electron transport scheme and the initial electron energy. The lowest cutoff energy allowed in the EGSnrc system is 1 keV, which is still a large value for the quantitative prediction of the $w$ results. Ideally one requires a method for the calculation of the electron fluence spectrum down to energies close to the ionization threshold but this is not possible with EGSnrc. One possible route to extend the electron differential fluence distribution to energies lower than the cutoff is through the use of the cavity integral in Eq. (11). This is based on the fact that the absorbed dose should not depend on the cutoff energy $\Delta$ [2]. Using Eq. (11), one can readily obtain the integral equation for the unknown fluence distribution

$$\int_{\Delta_1}^{E_0} \frac{d\Phi}{dE} L_m(\Delta_1, E) dE = \int_{\Delta_0}^{E_0} \frac{d\Phi}{dE} \big[ L_m(\Delta_0, E) - L_m(\Delta_1, E) \big] dE$$
$$+ \left[ \frac{d\Phi(\Delta_0)}{dE} S_m(\Delta_0) \Delta_0 - \frac{d\Phi(\Delta_1)}{dE} S_m(\Delta_1) \Delta_1 \right]$$

(13)

where $\Delta_0$ is the (1 keV) cutoff energy used in the calculations of the fluence distributions in Monte Carlo simulations and $\Delta_1$ is the new cutoff energy, which is lower than $\Delta_0$. Its value is determined from the following argument:

1.  The continuous-slowing-down approximation (CSDA) or the use of a stopping power concept for electrons in the energy range between $\Delta_1$ and $\Delta_0$ should still be a valid methodology in the treatment of energy losses by ionizing particles.

---

[2] This assumes implicitly that there is delta-ray equilibrium below any chosen value of $\Delta$; this will generally be the case to a very good approximation in a medium which is uniform over distances much greater than the range of electrons with energy equal to the maximum value of $\Delta$ involved, in this case 1 keV.

2.  It can be shown (Paretzke and Berger 1978) that the average fractional energy loss (i.e. relative to its kinetic energy) by an electron in a collision with a water molecule in vapor is 3.6 % at 1 keV, 6.4 % at 500 eV, and 22 % at 100 eV.

It is obvious that the use of a stopping power to describe the gradual energy loss along the electron track in air ceases to be meaningful at energies below few hundred eV. On the other hand, the collision stopping powers calculated directly from the experimental ionization and excitation cross sections show reasonable agreement with the Bethe formula (used in EGS4/EGSnrc Monte Carlo calculations) down to 200 eV (ICRU report 37). Therefore the requirement for the validity of the CSD approximation and correctness of the calculated collision stopping powers for air molecules have prompted us to set the new cutoff energy at $\Delta_1 = 200$ eV. Thus, the solution to the equation (13) will give us the unknown differential fluence distribution in the range between 200 eV and 1 keV (the fluence spectrum for electron energies 1 keV and upwards is known from the Monte Carlo simulations). We would also like to point out here that the 200 eV cutoff is below the K-shell binding energies for both oxygen (~540 eV) and nitrogen (~400 eV) atoms. The fact that it is below the K-shell binding energies for both species will undoubtedly introduce some correlation into the fluence energy spectrum. Combined with low-energy uncertainties stemming from the neglect of the atomic/molecular binding energies in the stopping power expressions, it is obvious that the electron fluence spectrum in the sub-500 eV energy range is somewhat compromised. At the same time, there have been many studies carried out with the EGS4 system, indicating that implications of violating the requirement for the cutoff to be larger than the binding energies of the medium are less severe than one might expect from purely theoretical arguments[18]

## 3. Results

The solution to equation (13) for the unknown portion of the fluence spectra was found using a numerical integration algorithm. Figure (4) shows the calculated electron differential fluence distributions in an air slab of thickness 0.5 cm located at different depths in a water phantom. A beam of monoenergetic electrons with initial energy of 10 MeV is incident on the phantom. The general feature of the fluence distributions below the energy of 1 keV is their gradual increase with decreasing electron energy. One can expect the electron fluence spectrum to increase even further as the energy decreases and to saturate somewhere around the ionization threshold. The exact functional dependence of the electron fluence spectrum below 200 eV would require a separate investigation, which should rely on low energy electron ionization and excitation (total and differential) cross sectional data.

Once the electron fluence spectra have been obtained, the $w_\Delta$ value can be readily calculated using equation (12). Figure (5) shows the $w$ dependence (calculated for air) on the penetration depth in water for an electron beam with initial energy 10 MeV. The $w$ value remains constant throughout the depths studied (0.75 cm $\leq d \leq$ 6.75 cm) with its absolute value $w \approx$ 41.4±0.2 eV. This is higher than the experimentally reported value of 34 eV (ICRU report 31 1979), which was expected as already mentioned due to our cutoff energy of 200 eV being appreciably greater than the ionization threshold.

The results shown in figure (5) confirm the widespread assumption of the constancy of the $w$ value for megavoltage electron beams, but disagree with earlier made hypothesis of more profound role of excitation collisions as well as earlier reported experimental measurements by Domen and Lamperti[35]. A more detailed discussion on the possible nature of this disagreement is given in the next section of the paper. It is well established (ICRU report 31 and references therein) that as the electron energy falls below a value of several keV range, the $w$ value exhibits a gradual increase and can reach the values as high as 1000 eV per ion pair formed for electrons with kinetic energies of 20 eV. Figure (6) shows the results of the present studies of $w$ dependence on the penetration depth in air for electrons with initial energies of 50 keV and 10 keV. It should be noted here that the depth regions in which the $w$ value was calculated did not extend beyond those for which the predictions of Monte Carlo depth-dose distributions (obtained from DOSRZnrc code) coincided with depth-dose distributions calculated using the cavity integral formulation, given by equation (11).

The absolute value of $w$ exhibits an increase with penetration depth only in regions where the average electron energy (obtained from the electron fluence spectrum with cutoff of 200 eV) is lower than 5 keV. Figure (7) shows the $w$ dependence on the penetration depth in air for electrons with initial energy of 50 keV and three different cutoffs. As one can see, higher cutoff energies result in a greater variation of $w$ with depth. With the 200 eV cutoff, however, there is virtually no variation in the $w$ value. This dependence on the cutoff will be discussed in more detail in the next section.

# 4. Discussion

In the previous section we have presented the results of computer simulation studies of the dependence of $w$ on the depth of the air cavity in water for a given initial electron energy. The main purpose of these studies is to test our initial conjecture that the contribution of excitation relative to ionization events changes in the megavoltage domain (due to relativistic effects) and consequently influences the absolute value of $w$ in this therapeutic energy range.

The subject of the $w$ value has been systematically studied by many authors. As we mentioned earlier Grosswendt and Waibel[12] had calculated $W$ values for non-relativistic initial electron energies using Monte Carlo simulations, directly sampling all the physically relevant processes without resorting to the condensed history and multiple scattering theories. The cross sectional data for inelastic collisions was taken from the analytical functions of Green and Stolarski,[36] with fitting parameters chosen to comply with experimental data of Kieffer and Dunn.[32] On the other hand, the analytical function fit of Green and Sawada[37] was used to evaluate the differential cross section $d\sigma/dk$ for ionization events. However, we have found that the integration of this differential cross section over the energies of the secondary electrons leads to total ionization cross sections that are not equal to those obtained from the functional fits of Green and Stolarski, which have been used in their simulations. This inconsistency will inevitably influence the calculations of the electron transport in the medium and introduce an error in the final computation of the $w$ value. Equally well the present work also relies heavily on the results of Monte Carlo simulations, albeit those that are based on the condensed-history approach to the solving the Boltzmänn transport equation.[38] Since Monte Carlo simulations based on the condensed-history simulation scheme and therefore involving multiple scattering theories represent an approximation to the solution of the charged-particle transport in the medium, one could argue that the results obtained from such simulations may contain an inherent error. Precise estimation of such an error is beyond the scope of the present article, but its influence on the calculation of the electron fluence spectra is of utmost importance for plausible predictions of the $w$ value. One of the sources of uncertainties pertained to the calculation of the unknown portion of the fluence distributions arises from the values of the stopping powers for low energy electrons (the Bethe stopping powers used in the EGSnrs system are obtained from combining the Bethe expression with that calculated using the Møller differential cross section, which only describes collisions between **free** electrons). Combined with the already mentioned restricted stopping-power uncertainty stemming from the low cutoff values employed in finding the unknown portion of fluence distributions, it is obvious that the calculated energy spectra in the low (sub-1 keV) energy range also involves these errors. In order to quantify the uncertainties introduced by using Bethe-Møller based stopping powers in the low-energy region, we decided to estimate how the variation in the stopping powers would influence the final results. An imposed variation of $\pm10\%$ on the stopping powers in the energy range between 200 eV and 1 keV revealed no significant change (an average variation of 0.5% for the absolute value of $w_\Delta$ was seen, but no functional dependence on energy was observed) in the calculated w/e dependence on energy for the megavoltage electrons. Since the main interest of this project was to investigate the w/e dependence on energy in the therapeutic range, this suggests that the use of the Bethe-Møller stopping powers for sub keV calculations is valid.

There are several points that we would like to make in order to justify the use of the condensed history Monte Carlo simulations in the calculation of $w$. First, the depth-dose distributions calculated using DOSRZ Monte Carlo code show a very good agreement with those measured experimentally.[6] The second point is related to the fact that the electron stopping power in air calculated using combined Bethe-Møller cross sections sensibly agrees with that calculated using experimentally measured ionization and excitation cross section data for electron energies down to 200 eV.[39] Because the condensed history Monte Carlo simulations completely rely on the electron stopping powers in the treatment of sub-threshold events in the calculation of the absorbed dose and the same dose can also be obtained from the cavity integral formulation, one is led to the conclusion that the differential electron fluence distributions calculated using the condensed-history and multiple-scattering approximations should be in a good agreement with those that can be measured experimentally.

One obvious question arising from these results is why $w$ remains constant throughout regions where the average energy of electrons is larger than several keV and exhibits an increase only in those regions where the average electron energy falls below this value. The variation of $w$ with energy/penetration depth can come from either the interplay between the excitation and ionization collisions or from the fact that low-energy electrons produce electron-ion pairs with spatial/energy distributions that lead to more initial electron-ion or ion-ion (an electron can attach itself to an oxygen molecule to form an oxygen ion) recombinations, thus *decreasing* the total number of ions collected. Since the electron attachment rate to the oxygen molecule is proportional to the concentration of neutral oxygen and the rate of electron recombination with ions is proportional to the concentration of ions, the electron attachment processes are overwhelmingly predominant over those of electron-ion recombination (concentration of neutrals is much higher than that of ions) leading only to a possibility of positive-negative ion recombinations.

The subject of the collection efficiency of air-filled ionization chambers has been well studied.[40] It can be stated that with the proper design, ion chambers should collect nearly all of the charges initially created by ionizing electrons or subsequently formed by attachment processes. This means that the interplay between excitation and ionization events is the only physical process relevant to changes in $w$.

The question that needs to be answered is whether the mere presence of excitation collisions could explain the observed dependence. Since excitation events are present at all electron energies and the $w$ value does not change at megavoltage electron energies, one is led to the conclusion that the different functional dependence of both the total excitation and ionization cross sections on the ionizing particle energy and/or the dependence of the mean excitation energy (related to energy losses due to impact excitation) on the ionizing electron energy will bring about variations in the $w$ value. Indeed if both, the total excitation and ionization cross sections have the same functional dependence on the electron energy i.e. $\sigma_{ex}(E)/\sigma_{in}(E) = $ const. (no matter what the absolute value of the ratio may be) and the same energy (mean energy of excitation) is transferred to the medium in excitation events irrespective of the electron energy then $w$ must be constant for all relevant electron energies (energies higher than the ionization threshold). In reality however, both the mean excitation energy and the ratio between the total excitation and ionization cross sections are functions of ionizing electron energy.

As shown by Grosswendt and Waibel,[12] the mean excitation energy for molecular nitrogen increases from 7 eV to 12.8 eV for primary electrons in the energy range 10 eV to 50 eV and then remains nearly constant for higher energies. An analytical model for electron excitation cross sections was proposed by Green and Stolarski[36] in which the authors found an analytical fit to experimentally measured electron excitation cross sections for molecular nitrogen and oxygen. Using this model we readily arrive at the total (summed over all the major excitation lines) excitation cross section for air. Figure (8) shows the ratio between the total excitation and ionization cross sections as a function of incident electron energy. As expected, this function exhibits dramatic variations (excitation events becoming dominant over ionization) only for electrons with energies lower than around 300 eV. For higher energies the ratio remains nearly constant.

As mentioned earlier, results presented in this work show that $w_\Delta$ starts increasing in regions where the average electron energy is on the order of 5 keV. At the same time figure (8) suggests that the ratio between the number of excitation and ionization events for electrons with energy higher than 300 eV remains almost constant. Why therefore does the observed variation in $w_\Delta$ starts manifesting itself at much higher energies than 300 eV? The answer lies in the relative contribution of these sub-300 eV electrons to the total number of ion pairs created and total energy deposited in the medium. Once the average energy (in a given volume) becomes low enough, the contribution to ion creation of electrons with energies between the lowest ionization threshold and a few hundred electron volts (i.e. the energy region where the most profound dependence of the excitation-to-ionization cross section ratio on the ionizing electron energy occurs – see figure 8) becomes non-negligible, eventually leading to a gradual increase in $(w/e)_{air}$ (more energy is deposited without corresponding creation of ion pairs). At higher values of the average electron energy, the relative weight of these "subthreshold" electrons decreases, becoming insignificant at megavoltage energies, which explains the constancy of $w_\Delta$ for megavoltage electrons and also at energies much lower than this. The absolute value of our calculated $w_\Delta$ suggests that electrons with energies below the cutoff (200 eV) contribute ~20 % to the integral in Eq. (5). The question arising from this observation is: Given that there are still significant numbers of ion pairs (~20%) created by electrons in the energy range where significant variation of the excitation-to-ionization cross section ratio takes place, why is their contribution such that $w_\Delta$ stays constant for high energy particles? We think that the answer to this puzzle can be obtained through the following arguments. The differential $w$ value can be written in the following form,

$$ w = \frac{N_i^{(1)}\overline{\varepsilon}_i^{(1)} + N_{ex}^{(1)}\overline{\varepsilon}_{ex}^{(1)} + N_i^{(2)}\overline{\varepsilon}_i^{(2)} + N_{ex}^{(2)}\overline{\varepsilon}_{ex}^{(2)}}{N_i^{(1)} + N_i^{(2)}} \quad (14) $$

where $N_{i,ex}^{(1,2)}$ denote the number of ionization/excitation events created by electrons with energy above/below $\Delta$, and $\varepsilon_{i,ex}^{(1,2)}$ denote the average energy transferred to the medium during ionization/excitation collisions by electrons with energy above/below $\Delta$.

The term in the numerator represents the total energy absorbed in the medium and that in the denominator, the total number of ion pairs created. The structure of the energy absorption term presented above is somewhat different from that using the cavity integral approach. This arises from the fact that the former does not have the requirement for energy transfer to be less than the preset value of $\Delta$ (as it must do in the cavity integral approach). In this respect the division of collisions into above/below $\Delta$ in the above expression is just a mathematical split of one expression into the sum of two. In the cavity integral approach however, the cutoff value plays a much more fundamental role than just a simple algebraic split. One artificially limits the 'continuously' transferred energy to values below $\Delta$. As a result, the track-end term must be added to the cavity integral to account for those primary particles that cross the $\Delta$-boundary. In an analogous manner, $N_i^{(2)}$ is the number of ion pairs that is missing in our calculation model ($N_i^{(2)}/N_i^{(1)} \sim 0.2$). When the average primary electron energy in the slab is high and the value of $\Delta$ is small, the average energy transferred to the medium per collision by the above-$\Delta$ electrons is significantly greater than the average energy transferred by electrons below $\Delta$ (firstly, energy losses cannot be greater than $\Delta/2$ and secondly, higher energy electrons can sample the whole shell structure of the atom, whereas low energy electrons can only interact with outer shell atomic electrons where the binding energy is low). In this case the product $N_i^{(2)}\varepsilon_i^{(2)}$ is much smaller than the product $N_i^{(1)}\varepsilon_i^{(1)}$ even though $N_i^{(2)}$ or $\varepsilon_i^{(2)}$ by themselves can be non-infinitesimal. For the same reasons, the product $N_{ex}^{(2)}\varepsilon_{ex}^{(2)}$ makes an insignificant contribution to the total energy absorbed. When the average electron energy is high, the excitation term $N_{ex}^{(1)}\varepsilon_{ex}^{(1)}$ should be proportional to the ionization term $N_i^{(1)}\varepsilon_i^{(1)}$ because of the constancy of excitation-to-ionization ratio (which determines the relative numbers of ionization and excitation collisions). The results of MC simulations and the numerical integration of Eq. (11) also demonstrate that the ratio of the track-end part to the total energy absorbed is constant in any 'slab' where the average energy is much higher than the cutoff value. Only when this condition breaks down does the ratio of the track-end term to the cavity integral start changing. Since the cavity integral formulation of the absorbed dose should give the same absorbed energy as the equation presented above, one is led to the conclusion that the ratio $N_i^{(2)}\varepsilon_i^{(2)}/N_i^{(1)}\varepsilon_i^{(1)}$=const. when the average primary electron energy in the air 'slab' is much higher than the cutoff value. The above condition is equivalent to the following, $N_i^{(2)}/N_i^{(1)}$=const. since $\varepsilon_i^{(2)}/\varepsilon_i^{(1)}$=const. for the given value of the cutoff and much larger value of the electron energy. Therefore, the constancy of $w_\Delta$ for high-energy electrons calculated in this manuscript can be explained through two conditions. The first stems from the already mentioned fact that the excitation-to-ionization cross section ratio is basically constant for electrons with energies above 300 eV or $N_{ex}^{(1)}/N_i^{(1)}$=const. The second is related to the condition $N_i^{(2)}/N_i^{(1)}$=const. (this relation can only hold true together with the first condition). The satisfaction of these two conditions not only ensures the constancy of $w_\Delta$ but also that of the total differential $w$ value as can be seen from Eq. (14). This is why we can claim with a high degree of confidence that once the primary electron energy is such that the calculated $w_\Delta$ is constant for the given small value of the cutoff, the total differential $w$ value should also be independent of the electron energy/depth (as long as the average energy in the 'slab' is much higher than the cutoff). As the electron energy drops, the cutoff value becomes comparable to the average electron energy and the above conditions that ensure the constancy of $w_\Delta$ break down, leading to its variation with depth/electron energy.

It should be noted here that $w_\Delta$ variation with depth cannot be unequivocally linked to the increased role of excitation events only. Even if there were no excitation collisions present in a system, $w_\Delta$ (but not the $w$ value) calculated in this manuscript would still change if the cutoff value was comparable to the average electron energy. This can be seen from the fact that as the energy of the electrons drops, $\varepsilon_i^{(1)}$ will start decreasing as well, bringing in the energy dependence in both ratios- $N_i^{(2)}\varepsilon_i^{(2)}/N_i^{(1)}\varepsilon_i^{(1)}\neq$const, $N_i^{(2)}/N_i^{(1)}\neq$const. This in turn will lead to the energy dependence of $w_\Delta$. Obviously this variation is not due to the excitation collisions, but rather an artifact introduced by splitting particles into two energy domains. This is why we had to set the cutoff in our calculations to the lowest value that the physical reasoning had allowed us to. This would ensure the minimization of the cutoff artifact introduced by employing the condensed history approximation. As shown in Fig. (7), different gradients in the $w_\Delta$ dependence on depth are possible depending on the value of the cutoff (larger values of the cutoff result in steeper functional dependencies). In order to show unambiguously that the observed increase is due to the effect introduced by the excitation collisions, one would require the differential energy fluence distributions all the way down to the ionization threshold. Unfortunately the exact functional form of the fluence spectra in the range $I_0 < E < 200$ eV cannot be obtained from condensed-history MC simulations; one would have to employ an analogue (interaction-by-interaction) MC scheme, coupled with a knowledge of the appropriate energy-loss cross sections (ionization and excitation separately), to obtain the presently unknown differential fluence distributions; this is beyond the scope of the present investigation.

Experimental as well as computer-simulation results (ICRU Report 31 and references therein) suggest that $W$ starts increasing as the initial electron energy falls below several keV range. The ratio of the excitation to the ionization cross section shown in figure 8 strongly indicates that this rise in $w$ is due to the interplay between these ionization and excitation events though the limitation of a 200 eV cutoff in our electron fluence computations prevents us from being able to make precise statements about both this rise in $w$ and its absolute magnitude  However, despite this limitation, our theoretical study does allow us to conclude that $w$ exhibits no variation throughout the entire megavoltage energy range (within the uncertainties in the ionization cross section dependence on the electron energy) thus disproving our initial hypothesis, based on classical-physics ideas of the increased role that the excitation collisions might play, relative to ionizations, in the interactions of *relativistic* electrons with air molecules.

# 5. Summary

A combined theoretical and computer simulation study of the average energy required to create a single electron-ion pair in air in electron beams has been carried out. Using both EGSnrc condensed history Monte Carlo simulations and the Spencer-Attix cavity integral formulation for the absorbed dose we were able to extend electron fluence spectrum calculations from the Monte-Carlo cutoff of 1 keV down to 200 eV. The distributions of electron fluence differential in energy thus obtained were used to calculate the $w$ value as a function of electron penetration depth. The results of our calculations show that the $w$ value exhibits no discernible depth or energy dependence for ionizing electrons in the megavoltage energy range, thus confirming the widely adopted assumption of constant $(w/e)_{air}$ applied to the conversion of ionization chamber readings to absorbed dose to water in megavoltage radiotherapy beams. However, our computational approach did not allow us to compute $d\Phi/dE$ all the way down to the ionization threshold (approximately 15 eV in air) which inevitably resulted in our absolute value for $(w/e)_{air}$, 41.4 eV, being significantly higher than the best experimental number of close to 34 eV. We have also found that $w$ increases in regions where the average electron energy is below 5 keV consistent with the experimental literature. However, because the minimum cutoff energy (200 eV) used in our calculations is still large compared to the ionization threshold, the observed increase in $(w/e)_{air}$ cannot be unequivocally linked to the increased role of the excitation events at these very low energies.

Attix K R Kase (Academic Press, New York, 1987), pp. 169-243.

# Figure captions

Fig.1 A schematic diagram of the simulation geometry. Z denotes the depth in the water of the air slab of thickness 0.5 cm.

Fig2. Total electron-impact ionization cross section of nitrogen molecule as a function of electron kinetic energy.

Fig3. Total electron-impact ionization cross section of oxygen molecule as a function of electron kinetic energy

Fig4. Electron differential fluence distribution at different depths in water. The initial electron energy is 10 MeV

Fig5. The dependence of the $w$ value on the penetration depth in water for the initial electron beam energy of 10 MeV. The global cutoff energy $\Delta=200$ eV.

Fig6. The dependence of the $w$ value on the penetration depth in air for the initial electron beam energies of 50 keV and 10 keV correspondingly. The global cutoff energy $\Delta=200$ eV

Fig7. The dependence of the $w$ value on the penetration depth in air for the initial electron beam energy of 50 keV for three different cutoffs. The solid line corresponds to $\Delta=200$ eV, the dotted line to $\Delta=500$ eV and the dashed-dotted line to $\Delta=1$ keV.

Fig8. The ratio of excitation to ionization cross section for air molecules (22 % $O_2$ and 78 % $N_2$) versus electron kinetic energy.

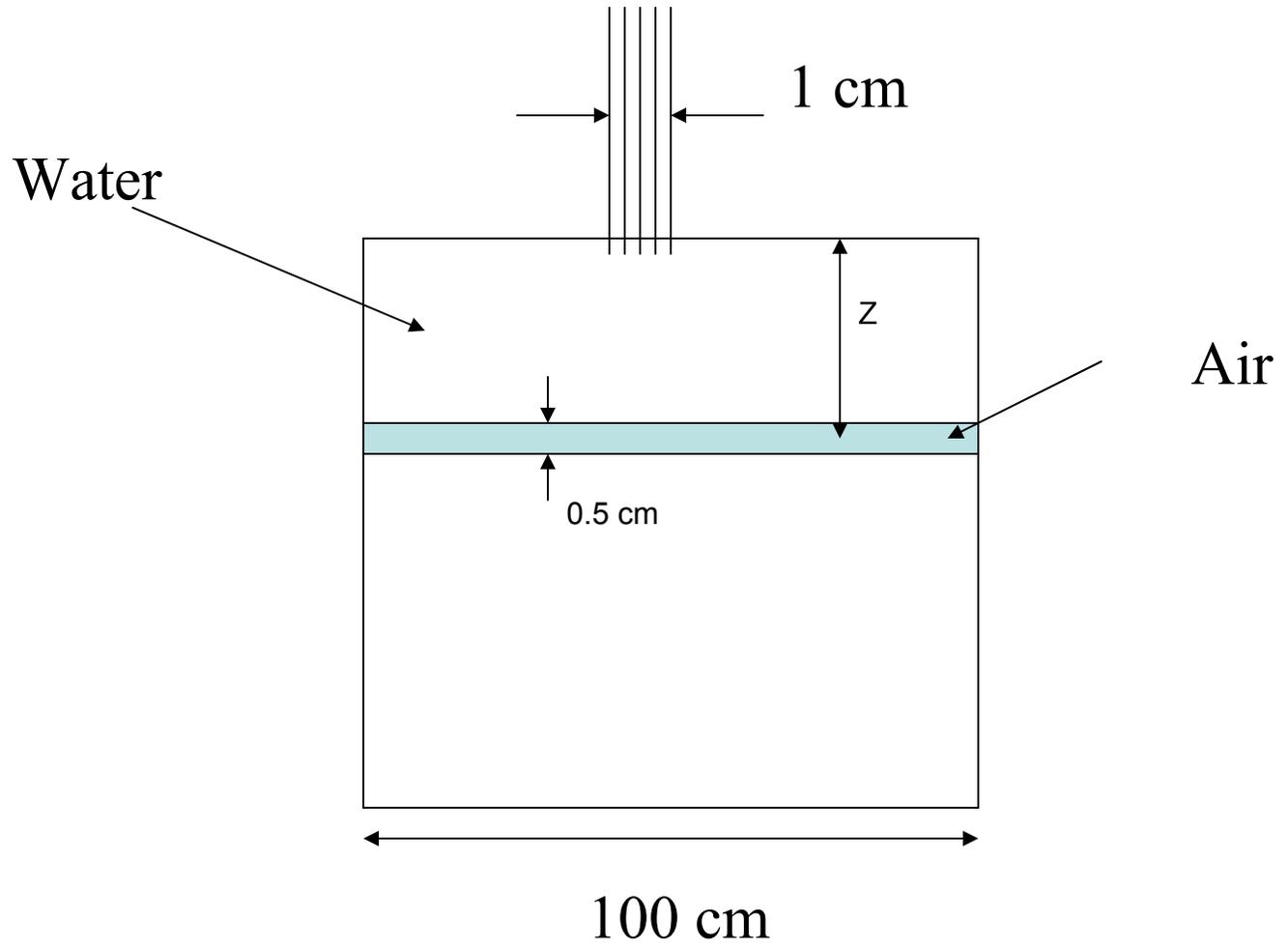

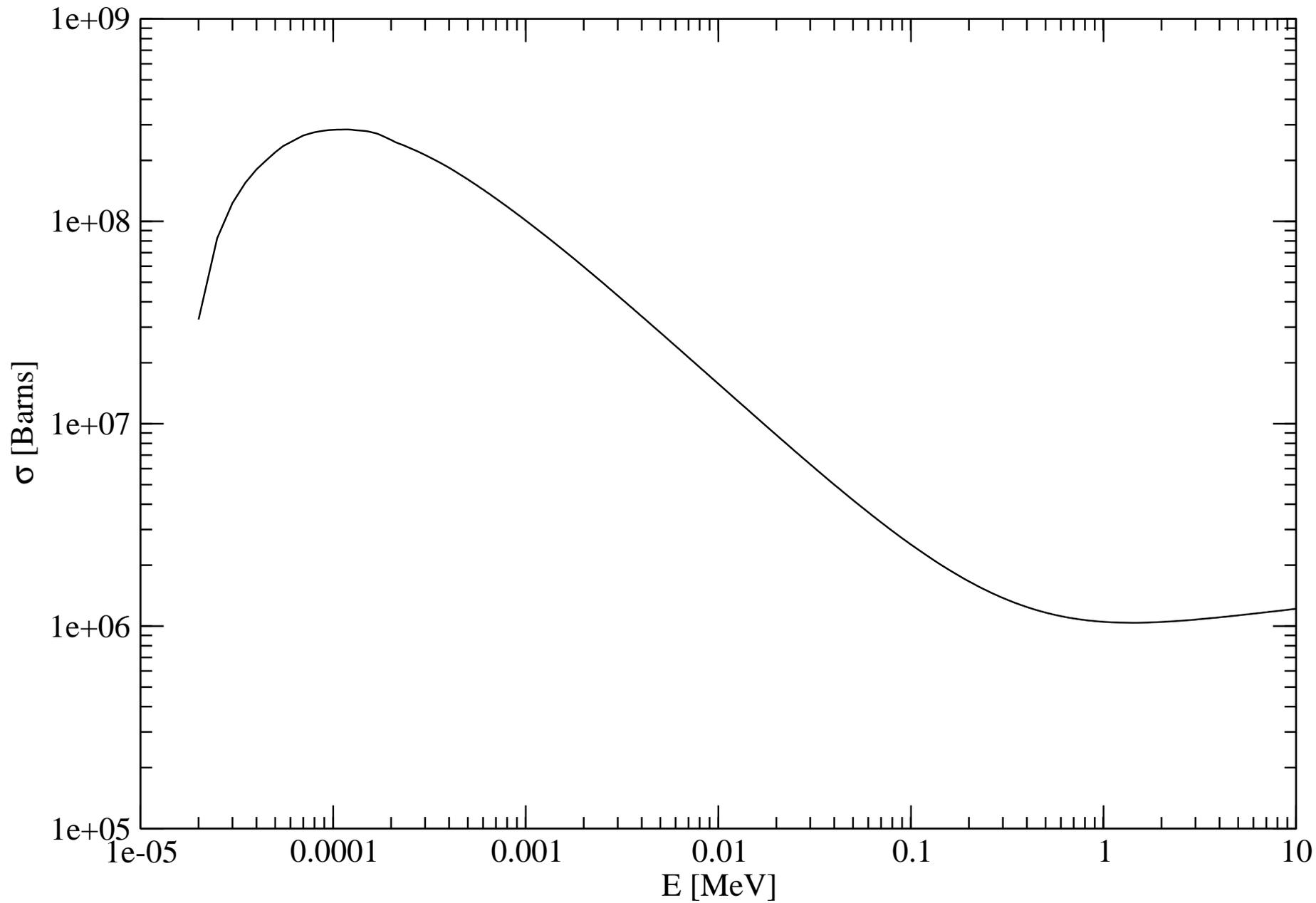

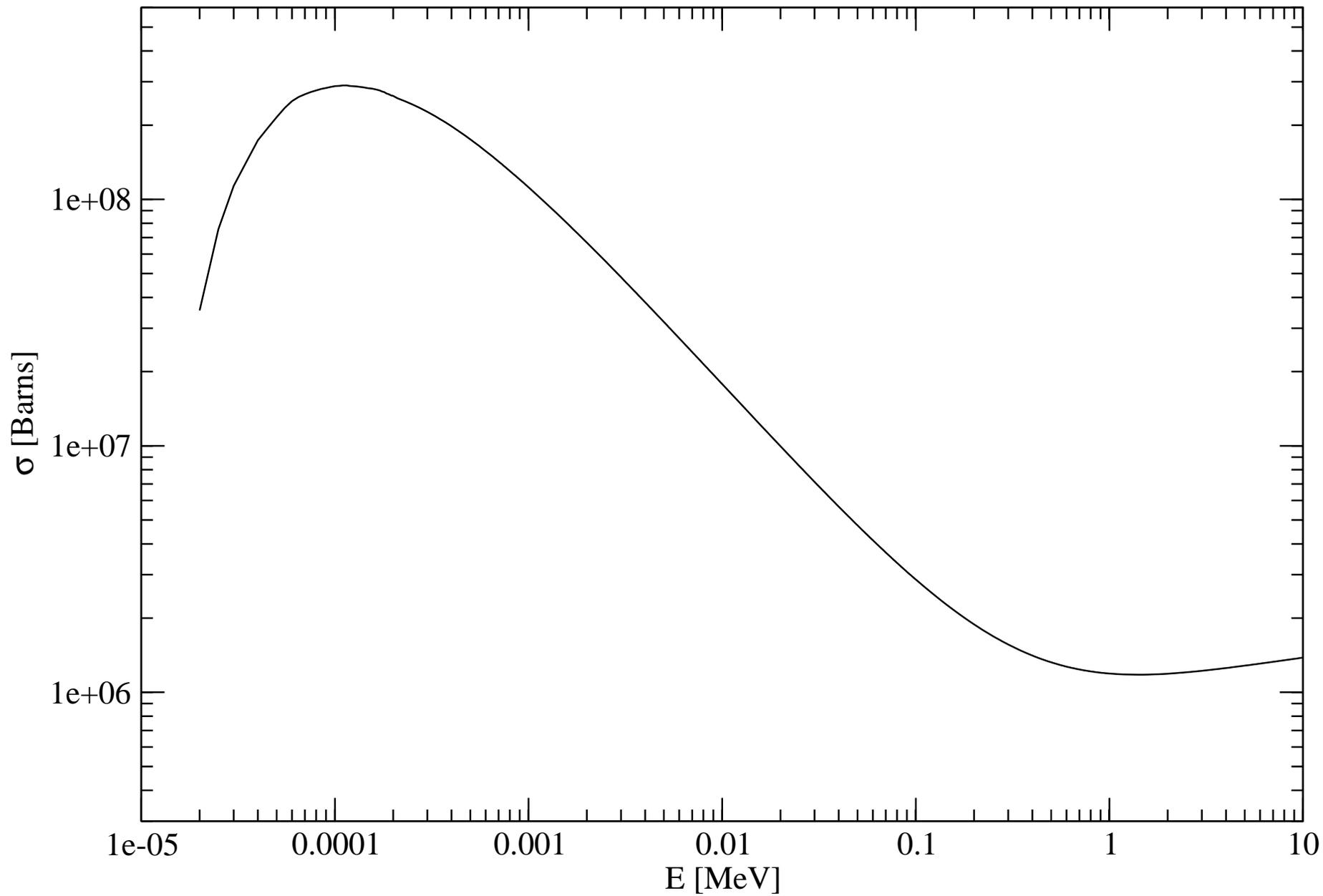

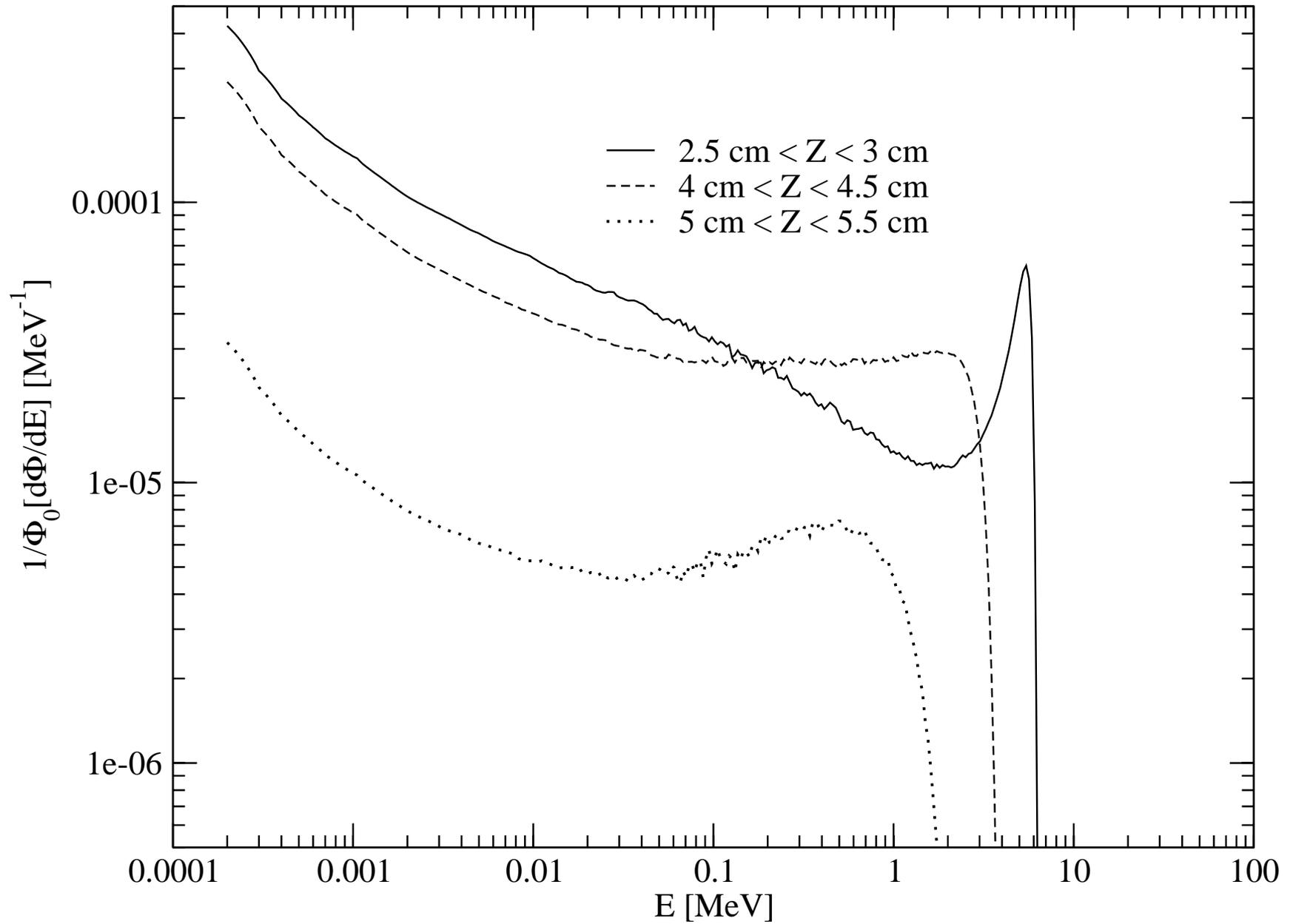

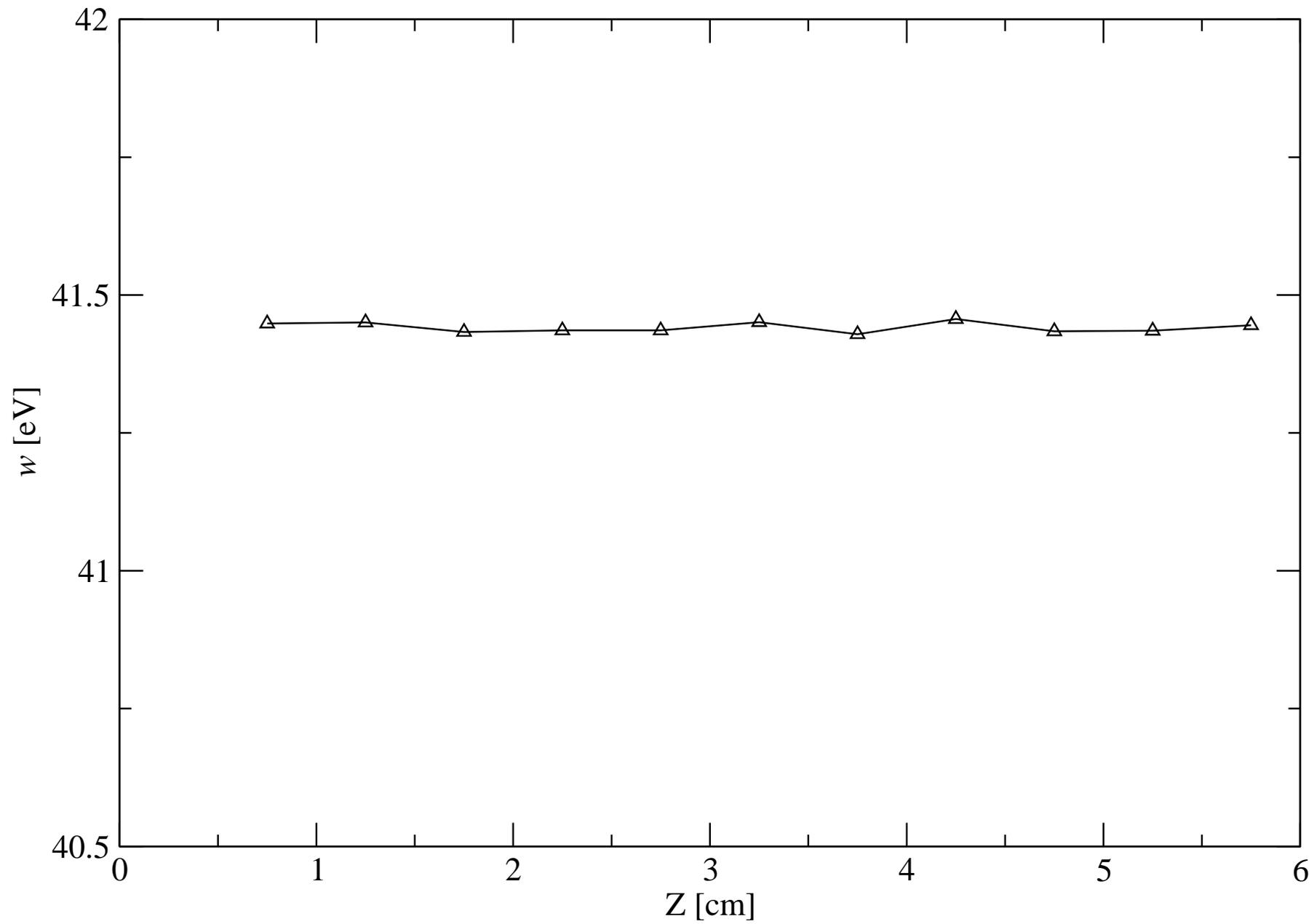

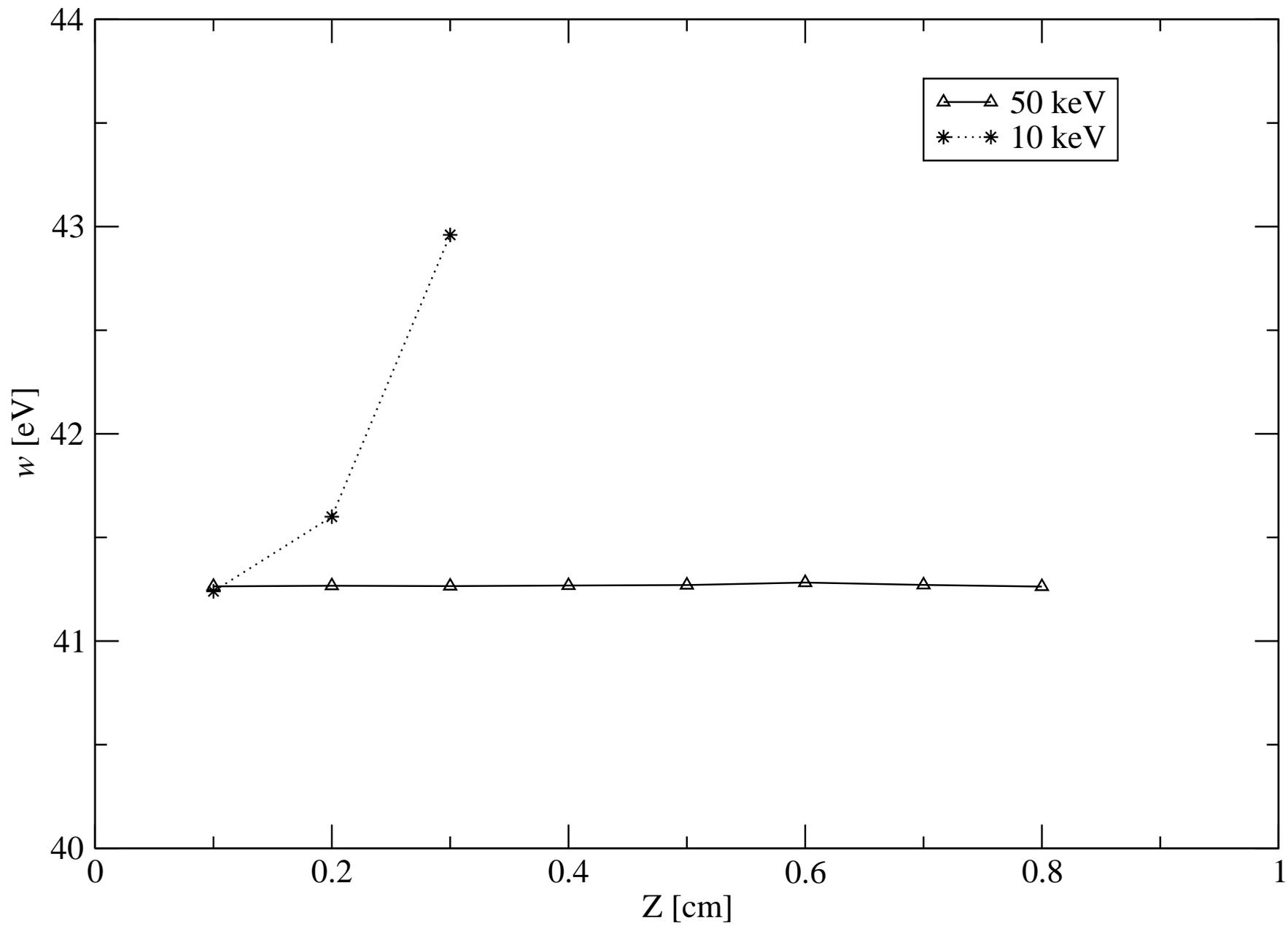

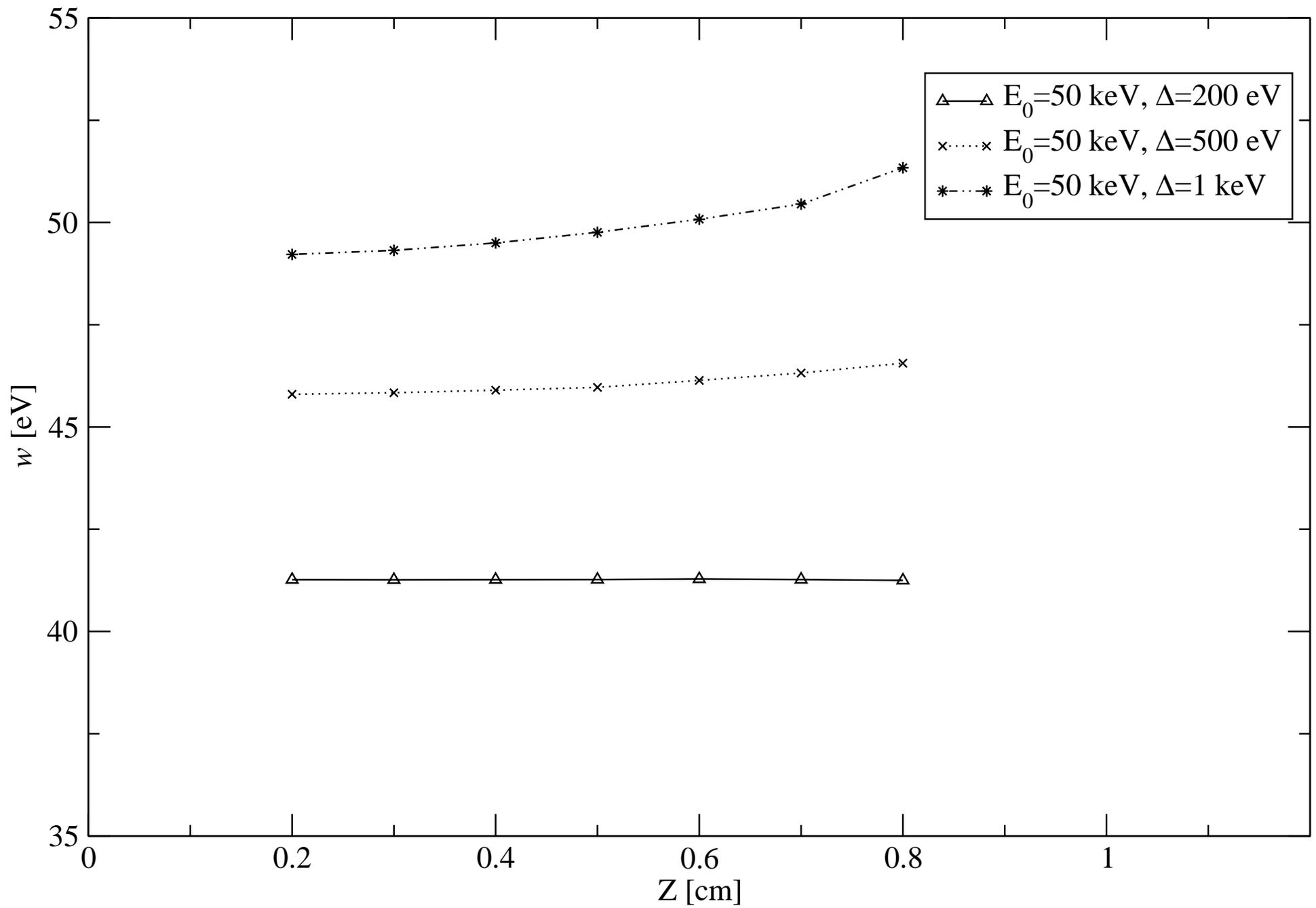

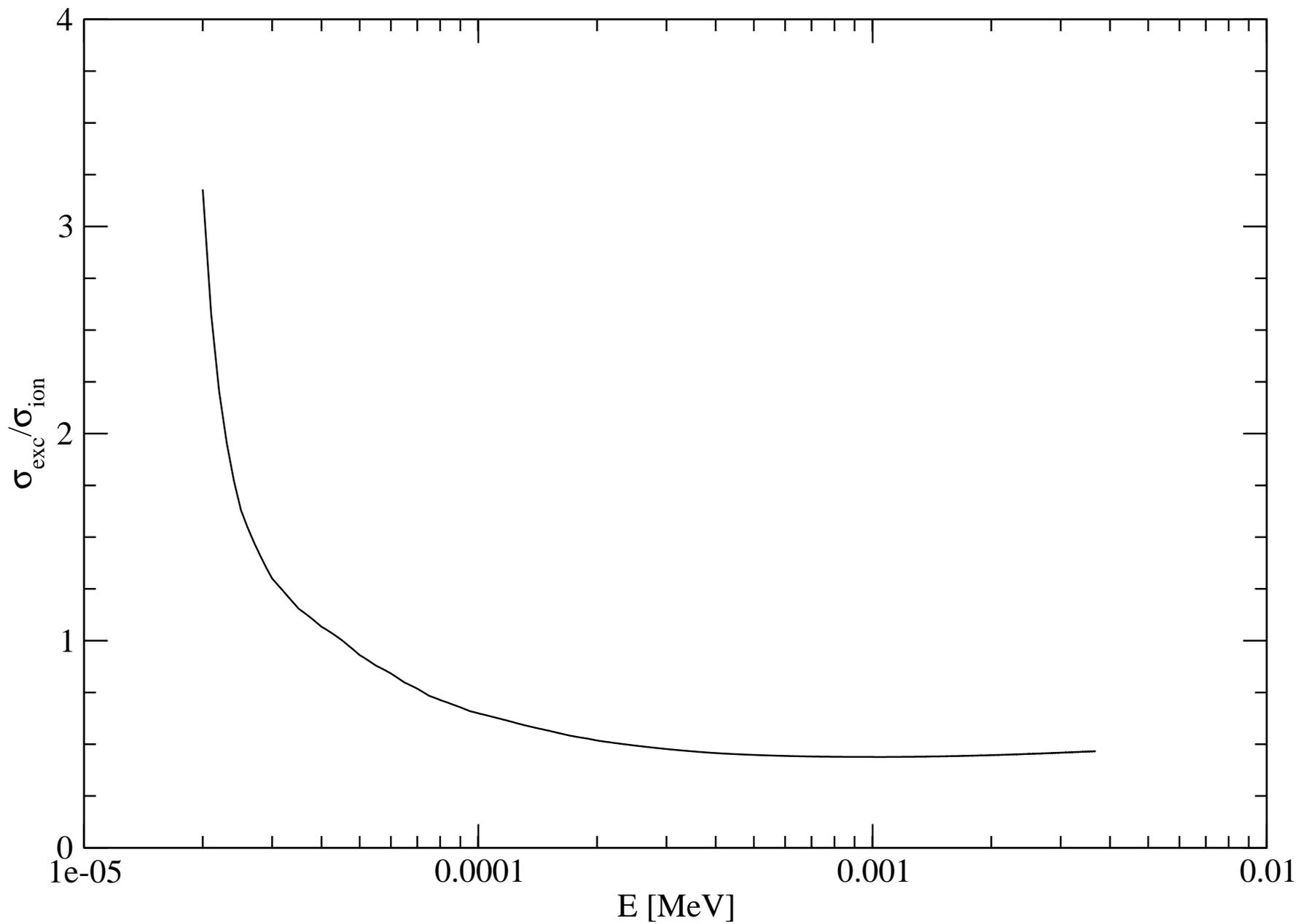